\documentclass[conference]{IEEEtran}

\usepackage{cite}
\usepackage{amsmath,amssymb,amsfonts}
\usepackage{graphicx}
\usepackage{textcomp}
\usepackage{xcolor}
\usepackage{url}
\usepackage{pgfplots}
\usepackage[protrusion=false]{microtype}
\usepgfplotslibrary{groupplots}
\pgfplotsset{compat=1.18}

\usepackage[hidelinks]{hyperref}

\title{Attack Injection to Vehicle Reactions: Extensible Closed-Loop Adversarial Evaluation Framework}

\author{
\IEEEauthorblockN{
Mohamed Gaber\IEEEauthorrefmark{1},
Tahsin Wahid\IEEEauthorrefmark{1},
Mohammad Mamun\IEEEauthorrefmark{2},
Salimur Choudhury\IEEEauthorrefmark{1}
}

\IEEEauthorblockA{
\IEEEauthorrefmark{1}School of Computing,
Queen's University, Kingston, Ontario, Canada
}

\IEEEauthorblockA{
\IEEEauthorrefmark{2}National Research Council Canada,
Fredericton, NB, Canada
}
Email: \{mohamed.gaber, tahsin.wahid, s.choudhury\}@queensu.ca, mohammad.mamun@nrc-cnrc.gc.ca
}

\begin{document}

\maketitle

\begin{abstract}

Autonomous vehicles can use information exchanged with other vehicles and roadside infrastructure to support driving decisions. However, applications that rely on this information can make unsafe decisions when it is falsified. Studying the resulting traffic effects requires following received information through application decisions, vehicle reactions, and subsequent interactions. We present an extensible closed-loop adversarial evaluation framework built on VaN3Twin, a platform for traffic and wireless network simulation, with a diverse attack library and a labelled dataset covering benign and attack conditions. A key advantage of the framework is that it can be extended with additional attacks, message paths, receiving applications, and logging fields. The added library contains 76 configurable attack profiles adapted to message formats defined by the European Telecommunications Standards Institute. These profiles are organized into nine groups of related attacks, referred to as suites. The framework connects live attack injection and simulated network delivery to receiving applications, whose selected commands update vehicle movement. The resulting state feeds subsequent message generation, closing the loop. The dataset links physical vehicle states and message exchanges to attack labels, application decisions, selected commands, and traffic outcomes. Evaluation with individual and combined attack suites confirmed execution of all profiles, and every suite produced requests linked to attacked messages. A recorded sequence also showed a vehicle reacting to an attacked message and subsequently influencing another vehicle through an unmodified message describing its updated movement. These findings demonstrate how live injection and vehicle feedback support the study of direct and indirect effects as traffic evolves.
\end{abstract}

\vspace{-0.3cm}
\section{Introduction}
\vspace{-0.1 cm}

Autonomous vehicles (AVs) have the potential to improve road safety by reducing crashes caused by human error~\cite{nhtsaAutomatedSafety}. They use sensors such as cameras, light detection and ranging (LiDAR), and radar to perceive their surroundings. However, poor weather or blocked views can make it harder to detect nearby vehicles and obstacles. Cooperative Intelligent Transport Systems (C-ITS) allow vehicles and roadside infrastructure to exchange data through vehicle-to-everything (V2X) communication~\cite{vasp2023}. This gives a vehicle access to information beyond what its own sensors can detect. The European Telecommunications Standards Institute (ETSI) specifies several message types for these exchanges in Europe, 
whereas SAE International standards define the corresponding message protocols for North America. Cooperative Awareness Messages (CAMs) report the sender's position, speed, and heading~\cite{etsi2026cam}. Collective Perception Messages (CPMs) share information about objects detected by the sender's sensors, such as nearby vehicles and pedestrians~\cite{etsi2023cpm}. Decentralized Environmental Notification Messages (DENMs) report road hazards and accidents~\cite{etsi2025denm}.



However, a vehicle that relies on received information may make an unsafe driving decision if that information is false. Two examples illustrate the possible consequences of falsified CAM reports. In the first, a leading vehicle reports a false speed or position, so the following vehicle may not slow down in time and could collide with it. In the second, false reports from a vehicle approaching an intersection can cause the receiver to underestimate collision risk, potentially leading to a side-impact (T-bone) collision. Either outcome can cause severe injury. Both scenarios show why the accuracy of received information matters for road safety. Message authentication alone does not prevent these attacks. Cryptographic protection can verify a message's origin and integrity, but does not guarantee that the information it carries is true~\cite{vasp2023}. An authorized but malicious sender can still report false positions, speeds, accelerations, or events, or reuse outdated information.

These threats have motivated research on misbehaviour detection and the datasets needed to compare detection methods. VeReMi introduced a public, labelled dataset for this purpose, initially covering five position-based attacks~\cite{veremi2018}. F2MD supports a wider range of simulated attacks, together with misbehaviour detection and reporting~\cite{f2md2020}. VASP provides 68 attacks on Basic Safety Messages (BSMs), the IEEE/SAE message used in North America that is equivalent to ETSI's CAM, including attacks targeting emergency-braking and intersection-assistance applications~\cite{vasp2023}. It also considers how these applications respond to false information. Simutack demonstrates physical consequences of sensor and communication attacks, with its V2X example using a three-vehicle platoon~\cite{simutack2023}. Within ETSI-based systems, collective-perception research has also connected CPM attacks and mitigation to vehicle reactions, including braking at intersections~\cite{systemx2024}. 
Other studies have examined DENM misbehaviour and its implications for standardization~\cite{denmmisbehavior2024}.
These studies differ in the attacks they support, the messages they use, the effects they evaluate, and the range of information recorded for evaluation. Our work brings diverse attack behaviours into a common, ETSI-aligned environment, with logging that tracks their effects from message reception through vehicle reactions and subsequent vehicle-to-vehicle interactions in the traffic simulation.

To address these challenges, we extend VaN3Twin~\cite{van3twin}, a traffic and wireless network simulation platform with an attack library and closed-loop vehicle reactions. The framework connects attack injection and network delivery to application decisions, vehicle movement, and traceable outcome logging. Our objective is to study how vehicles react to attacks and how their reactions affect other vehicles within a common simulation environment. Our main contributions are:

\begin{itemize}
\item \textbf{A diverse attack library:}\footnote{Public dataset: \url{https://doi.org/10.4224/40004118}} we design a library of 76 diverse, configurable attack profiles. They cover reported vehicle states, message timing, replay, false identities, ghost vehicles, and attacks targeting specific road-safety applications.

\item \textbf{An extensible closed-loop adversarial evaluation framework:} the framework connects traffic and network simulation, attack injection, receiving applications, action arbitration, vehicle reactions, and outcome logging. Its design supports additional attack profiles, V2X message paths, receiver applications, and logging fields.

\item \textbf{A labelled automotive evaluation dataset:}\footnotemark[1] for benign and attack cases, it records experiment settings, physical vehicle states, transmitted and received CAM, CPM, and DENM messages, ground-truth attack labels, application decisions and outputs, reaction requests, arbitration decisions, selected speed commands, vehicle trajectories, encounters, surrogate-safety measures, and collisions.
\end{itemize}

\vspace{-0.2 cm}
\section{Related Work}
\vspace{-0.1 cm}



V2X message attacks have motivated research on misbehaviour detection and datasets for comparing detection methods. VeReMi introduced a public, labelled dataset, providing simulated logs with ground-truth labels for five position-falsification BSM attacks~\cite{veremi2018}. These attacks are generated during simulation, while detectors are evaluated afterwards using the recorded messages. VeReMi NextGen extends this work with 15 CAM attacks, multiple driver profiles, urban and highway scenarios, and predefined training, validation, and test splits~\cite{vereminextgen2026}. Unlike the original VeReMi, it applies attacks to the recorded messages in post-processing so that new attacks can be added without rerunning the simulation. These datasets enable reproducible detection experiments. However, their workflow does not feed receiver decisions back into the traffic simulation, so they do not show how reactions to attacked messages affect subsequent vehicle movement.


Other studies provide frameworks for generating attacks and evaluating detection during simulation. F2MD combines simulated misbehaviour with local detection, reporting to a misbehaviour authority, and dataset generation~\cite{f2md2020}. However, its vehicle-message attacks are limited to position and speed falsification, replay, flooding, and Sybil attacks, while its evaluation focuses on detection and reporting rather than vehicle movement. VASP expands the attack library to include attacks targeting Emergency Electronic Brake Light (EEBL) and Intersection Movement Assist (IMA) applications~\cite{vasp2023}. It illustrates how these attacks can influence receiving vehicles, while its evaluation examines detection performance. Although the attack library was broadened, the attacks target SAE BSMs. To support the ETSI standard, A4MD adapts F2MD attacks and detectors to ETSI CAMs and supports reporting suspicious messages to a central authority for validation~\cite{a4md2023}.


Another line of work examines vehicle reactions in collective-perception systems. MISO-V uses observations shared by neighbouring vehicles to detect a ghost vehicle introduced through falsified BSMs~\cite{misov2021}. Its T-junction scenario shows how rejecting the ghost changes the decision of the tested vehicle. Zhang et al. integrate CPM attacks, perception fusion, detection, and mitigation in an Artery-based framework that also incorporates F2MD functionality for CAMs~\cite{systemx2024}. Their intersection-assistance application can trigger Simulation of Urban MObility (SUMO) braking through the Traffic Control Interface (TraCI), and they provide datasets for extended perception and misbehaviour reporting. These works close the loop between a falsified message and the vehicle's reaction. However, their reported evaluations focus on collective-perception security: MISO-V examines a ghost-vehicle scenario, while Zhang et al. evaluate six CPM position and speed attacks under two target-selection modes. Their reported vehicle-reaction examples concern driving at intersections rather than comparisons across multiple receiving safety applications.


Some frameworks also consider attacks on both communication and sensor inputs. Simutack provides closed-loop vehicle control to demonstrate the effects of Global Navigation Satellite System (GNSS), camera-blinding, and V2X message-falsification attacks~\cite{simutack2023}. Anagnostopoulos et al. combine CAM manipulation, LiDAR and image attacks, and Global Positioning System (GPS) data spoofing within a common simulation environment~\cite{anagnostopoulos2026}. Their evaluation examines perception-model robustness and demonstrates how spoofed positioning affects cooperative localization and teleoperation. Although these frameworks cover multiple attack surfaces, their reported evaluations do not compare the simulated traffic effects of a broad library of ETSI message attacks.



Our framework brings a diverse, configurable attack library into an extensible ETSI-aligned environment with multiple receiving applications. The labelled dataset records physical vehicle states alongside transmitted and received messages, application decisions, selected commands, and traffic outcomes. These records allow reported information to be compared with physical ground truth and vehicle reactions to be traced back to the messages used in their decisions. This supports the study of both a vehicle's response to attacked information and how that response affects the subsequent movement of other vehicles.

\vspace{-0.2 cm}
\section{Framework Design}
\vspace{-0.1 cm}

\subsection{Background}
\vspace{-0.1 cm}

This subsection introduces the basic concepts on which our framework relies before presenting its design. VaN3Twin provides the simulation environment and ETSI message services used in our framework~\cite{van3twin}. SUMO models the road network and the movement of individual vehicles. Vehicle positions, speeds, and accelerations are updated as the traffic simulation advances. Wireless communication is modelled in ns-3, a discrete-event network simulator that handles packet transmission and reception. TraCI allows external applications to interact with a running SUMO simulation, both to read vehicle states and to issue commands that change vehicle movement. At the receiving vehicle, safety applications use the received information to assess traffic risks. The EEBL application warns following vehicles about emergency braking. ETSI includes this use case within Road Hazard Signalling, using DENMs~\cite{etsi2013rhs}. Intersection Collision Risk Warning (ICRW) uses information about vehicle movement and the intersection to assess possible collisions between vehicles and warn the driver when necessary~\cite{etsi2018icrw}.

\subsection{Architecture}
\vspace{-0.1 cm}

\begin{figure}[t]
\centering
\includegraphics[scale=0.4]{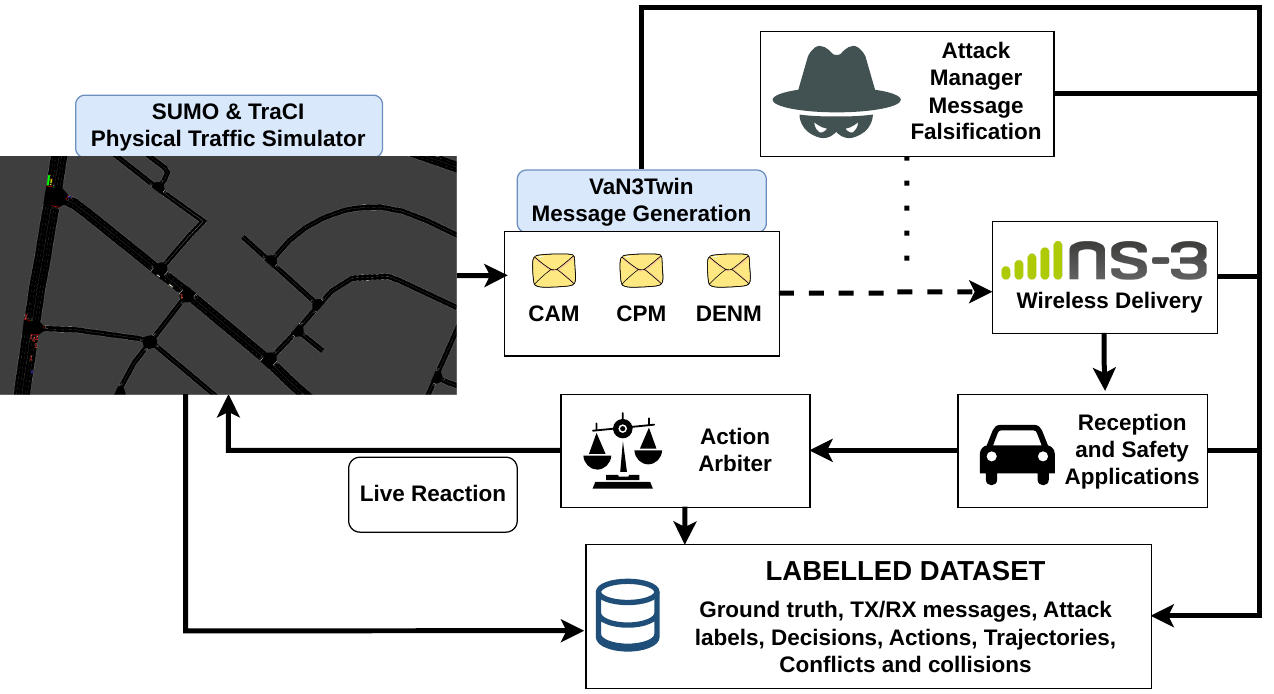}
\caption{The proposed closed-loop adversarial evaluation framework.}
\label{fig:framework-overview}
\vspace{-0.3cm}
\end{figure}

We extend VaN3Twin to study how attacks on exchanged messages affect vehicle decisions, movement, and subsequent traffic interactions. Fig.~\ref{fig:framework-overview} illustrates how attack generation, network delivery, receiving applications, and vehicle reactions are connected, with dataset logging throughout the process. VaN3Twin obtains physical vehicle states from SUMO through TraCI and uses them during V2X message generation. The Attack Manager can modify CAM information, change the CAM generation rate, or suppress CAM generation before encoding. DENMs are generated for application events, including fabricated emergency-braking events defined by attack profiles. Once encoded, the messages pass through the simulated wireless network and are decoded at receiving stations. At selected receiving vehicles, CAM and DENM information is processed by the vehicle applications. When an application requests a change in speed, its request passes through the action arbiter, which selects the command to be applied through TraCI. SUMO then updates vehicle movement, and the resulting state is used in subsequent messages, completing the feedback loop. The framework records physical vehicle states, transmitted and received messages, attack labels, application decisions, selected commands, and traffic outcomes. In Fig.~\ref{fig:framework-overview}, solid arrows represent data and control flow. The dashed arrow shows the outgoing message path, while the dotted line marks attack injection before transmission.



\vspace{-0.1 cm}

Our additions are organized into three parts: (1) an Attack Manager for configurable attack generation, (2) a vehicle-reaction layer that connects received information to vehicle movement, and (3) a dataset logging component to study the resulting vehicle decisions and physical traffic outcomes.

\subsection{Attack Generation}
\vspace{-0.1 cm}

The first addition is the Attack Manager, which introduces configurable adversarial behaviour into the simulation. Before each run, the Attack Manager loads and validates an attack plan containing attacker assignments, attack profiles, activation periods, and parameters. Each attacker corresponds to a vehicle participating in the simulated traffic and communication environment. The plan keeps these settings explicit and repeatable across runs. During its activation period, the assigned profile is applied when its required conditions are met. Depending on the profile, an attack may modify CAM fields, change the CAM generation rate, suppress CAM generation, reuse previously received information, or generate additional CAMs under false identities. Ghost-vehicle profiles report a fabricated vehicle without adding a physical vehicle to SUMO. False emergency-braking profiles also generate DENMs describing fabricated events. These messages pass through the attacker's simulated radio and follow the same network path as normal messages. We group the 76 profiles into nine suites according to the information they alter, their attack mechanisms, and the receiving applications they target, as summarized in Table~\ref{tab:attack-suites}. We also map the implemented attack behaviours to techniques in the Auto-ISAC Automotive Threat Matrix (ATM)~\cite{autoisacATM}. Table~\ref{tab:atm-mapping} assigns one primary mapping to each profile, while supporting mappings describe additional behaviours used by some profiles.

\begin{table}[t]
\caption{Implemented attack suites}
\vspace{-0.2cm}
\label{tab:attack-suites}
\centering
\footnotesize
\setlength{\tabcolsep}{2pt}
\begin{tabular}{@{}ll@{}}
\hline
\textbf{Suite (no. of profiles)} & \textbf{Information or behaviour} \\
\hline
Position and track presence (5) & CAM position and suppression \\
Speed and longitudinal motion (7) & CAM speed and motion \\
Acceleration (6) & CAM longitudinal acceleration \\
Heading and yaw rate (19) & CAM heading and yaw rate \\
Vehicle dimensions (18) & CAM length and width \\
Replay and stale data (3) & Replayed/stale CAM state \\
Flooding and message rate (3) & CAM rate and state \\
False identity, Sybil, and ghost vehicles (7) & CAM identity and ghost tracks \\
Application-targeted safety attacks (8) & EEBL/ICRW CAMs, DENMs \\
\hline
\textbf{Total (76)} & \\
\hline
\end{tabular}
\vspace{-0.3cm}
\end{table}

\vspace{-0.3cm}

\begin{table}[t]
  \caption{Mapping of attack profiles to the ATM}
  \vspace{-0.2cm}
  \label{tab:atm-mapping}
  \centering
  \footnotesize
  \setlength{\tabcolsep}{3pt}
  \begin{tabular}{c c c p{0.36\columnwidth}}
  \hline
  \textbf{Tactic} & \textbf{Technique} & \textbf{Profiles} & \textbf{Brief mapping} \\
  \hline
  ATM-TA0001 & ATM-T0003 & 71 & False, replayed, suppressed, or fabricated message information \\
  ATM-TA0001 & ATM-T0002 & 5 & High-rate message flooding \\
  ATM-TA0010 & ATM-T0038 & 11* & Collection of received vehicle information \\
  \hline
  \end{tabular}

  \vspace{1mm}
  \raggedright
  \scriptsize *Supporting mapping that overlaps with the primary profile counts.
  \vspace{-0.4cm}
  \end{table}

\vspace{4pt}
\subsection{Vehicle Reaction}
\vspace{-0.1 cm}
Receiving an attacked message does not, by itself, show whether the attack changes vehicle movement. We therefore add a vehicle-reaction layer that allows selected vehicles in SUMO to react to received messages. This layer contains three applications. The CAM-following application uses the reported position and motion of a vehicle that appears to be ahead to determine a target speed. EEBL evaluates emergency-braking events reported in DENMs and can request braking when an event is relevant to the receiving vehicle. ICRW uses reported vehicle motion and the road layout to identify possible conflicts at intersections. More than one application may request a reaction at the same time, so their requests pass through a common action arbiter. In this framework, EEBL requests have the highest priority, followed by ICRW and CAM-following requests. If multiple requests have the same priority, the lower target speed is selected. Each request remains active for a limited time and is removed when it expires or is withdrawn. The arbiter applies the selected speed command to the vehicle through TraCI. The resulting vehicle movement affects later simulation states and subsequent message exchanges, completing the feedback loop.

\subsection{Dataset Construction}
\vspace{-0.1 cm}

To study these reactions, the framework records messages, application decisions, selected commands, and traffic outcomes alongside physical vehicle states. Reported message values are kept separate from the physical vehicle state so they can be compared with ground truth. Transmission and reception records share an internal identifier, allowing a message to be traced across the communication path. Table~\ref{tab:dataset-output} summarizes the data recorded in both benign and attack cases.









\begin{table}[t]
\caption{Data recorded by the framework}
\vspace{-0.2cm}
\label{tab:dataset-output}
\centering
\footnotesize
\setlength{\tabcolsep}{3pt}
\begin{tabular}{@{}p{0.24\columnwidth}p{0.72\columnwidth}@{}}
\hline
\raggedright\textbf{Output group}
& \raggedright\textbf{Recorded information} \tabularnewline
\hline

\raggedright Vehicle state and movement
& \raggedright Physical position, speed, acceleration, heading,
yaw rate, dimensions, lane, and trajectories. \tabularnewline

\raggedright Communication
& \raggedright CAM, CPM, and DENM transmissions and receptions;
physical and advertised identities; timing, attack labels,
radio measurements, pre-attack and transmitted CAM values,
and suppressed CAM opportunities. \tabularnewline

\raggedright Applications and reactions
& \raggedright CAM-following, EEBL, and ICRW decisions;
application-generated messages; reaction requests and expiry;
arbitration results, selected speed commands, and ICRW
encounters. \tabularnewline

\raggedright Safety outcomes
& \raggedright Collisions; time to collision, deceleration rate
to avoid a crash, and post-encroachment time. \tabularnewline

\raggedright Experiment description
& \raggedright Scenario settings, seeds, routes, attacker and
controller assignments, attack parameters, message configuration,
schema, units, and completion state. \tabularnewline

\hline
\end{tabular}
\vspace{-0.3 cm}
\end{table}

\vspace{-0.2 cm}
\section{Experimental Setup \& Methodology}
\vspace{-0.1 cm}






In this section, we present the experimental setup for our proposed closed-loop framework. We evaluate the framework under benign and attack conditions to examine how injected attacks influence vehicle reactions and subsequent traffic interactions.
 \vspace{-0.2 cm}
\subsection{Simulation Scenario and Configuration}
\vspace{-0.1 cm}

We evaluate the framework using a traffic scenario derived from the LuSTNano configuration distributed with F2MD~\cite{f2md2020}. The traffic input contains 1,870 vehicles scheduled to depart during the first 3,000 simulated seconds. Each case runs for 3,600 simulated seconds, allowing traffic to continue for 600 seconds after the scheduled departure window. Table~\ref{tab:experimental-configuration} summarizes the settings used in these experiments.

\begin{table}[t]
\caption{Experimental configuration}
\vspace{-0.2cm}
\label{tab:experimental-configuration}
\centering
\footnotesize
\begin{tabular}{p{0.40\columnwidth} p{0.49\columnwidth}}
\hline
\textbf{Setting} & \textbf{Value} \\
\hline
Case duration & 3,600 s \\
Vehicles & 1,870 \\
Traffic simulation step & 0.1 s \\
Nominal CAM interval & 1 s \\
Radio configuration & IEEE~802.11p, 10~MHz, 6~Mbit/s, 13~dBm \\
Message services & CAM, CPM, and DENM \\
Attack cases & 374 attackers (20\%) \\
Reaction-controlled vehicles & 150 vehicles (10\% of non-attackers) \\
Message security & Disabled \\
\hline
\end{tabular}
\vspace{-0.3 cm}
\end{table}








\vspace{-0.1 cm}
\subsection{Experiment Design}
\vspace{-0.1 cm}

The evaluation contains eleven cases: one benign case, nine cases enabling the attack suites in Table~\ref{tab:attack-suites} separately, and one combining all nine suites. The traffic routes, vehicle-to-station mapping, designated attacker population, controlled vehicles, network and application settings, and random seeds are kept fixed across cases. Vehicle reaction is enabled in every case. The attacker and controlled-vehicle populations do not overlap. Attack profiles are assigned deterministically and distributed as evenly as their requirements allow. Some profiles require specific opportunities. For example, an intersection attack requires the attacking vehicle to approach an intersection during the simulation. These opportunities are identified from the benign case. Target-dependent profiles use actual CAM receptions, while intersection attacks use junctions approached by the corresponding attacker during the benign run. For reaction-controlled vehicles, SUMO's car-following safe-speed check is disabled while application commands are active.

\vspace{-0.1 cm}
\subsection{Evaluation Metrics}
\vspace{-0.1 cm}

We check attack execution using records where an attack was applied, rather than counting profile assignments. Delivery is verified by matching decoded reception records to their corresponding transmissions. Suppressed CAM opportunities are counted separately because no packet is generated. Attacker transmissions include all logged messages sent by designated attacker vehicles, including unchanged messages. We then count selected attack-fed requests: speed requests chosen by the action arbiter and linked to a received message with an applied attack. Each submitted or refreshed request is counted once if selected while the controlled vehicle is present. ICRW requests are included only when there is also a current warning for the same intersection encounter. We also report the number of distinct controlled vehicles receiving these commands, counting each vehicle once per case. Traffic outcomes are examined through vehicle trajectories and collision events. We report total collisions and their change relative to the benign case. A collision is classified as attack-associated when either participant has an active, selected attack-fed command at the time of the collision. Each collision is counted once, even if both participants qualify. This association does not prove causation. Alongside collisions, we examine time to collision (TTC), which estimates the time remaining before a collision under SUMO's motion assumptions~\cite{sumoSSM}. These values are calculated from physical vehicle states. We combine overlapping records for the same two vehicles and retain the smallest valid TTC. Each encounter is counted once if this minimum is below 1.5~s, using the threshold documented in~\cite{fhwaSSAM2008}.


\vspace{-0.1 cm}
\section{Results \& Discussion}
\vspace{-0.1 cm}

\subsection{Attack Execution and Network Delivery}
\vspace{-0.1 cm}

We first verified attack execution and delivery through ns-3. All 76 profiles executed in their respective suite runs and in the combined run. Sudden disappearance was verified through suppressed CAM generation. Across experiments, messages associated with applied attacks reached other vehicles for all 75 packet-producing profiles. Across all eleven cases, every reception record matched a corresponding transmission record.

\subsection{Application Responses and Direct Vehicle Reactions}
\vspace{-0.1 cm}

\begin{figure}[t]
\centering
\definecolor{ttcblue}{HTML}{1261A0}
\begin{tikzpicture}
\begin{axis}[
  xbar,
  scale only axis,
  width=0.62\columnwidth,
  height=0.57\columnwidth,
  xmin=0, xmax=1070,
  ymin=0.45, ymax=11.55,
  xtick={0,250,500,750,1000},
  ytick={1,2,3,4,5,6,7,8,9,10,11},
  yticklabels={All suites combined,Application safety,Identity/Sybil/ghost,Flooding/rate,Replay/stale,Vehicle dimensions,Heading/yaw,Acceleration,Speed/motion,Position/track,Benign},
  xlabel={Encounters with minimum TTC $<1.5$ s},
  tick label style={font=\scriptsize},
  label style={font=\scriptsize},
  scaled x ticks=false,
  x tick label style={/pgf/number format/fixed,/pgf/number format/1000 sep={}},
  axis x line*=bottom,
  axis y line*=left,
  y axis line style={draw=none},
  ytick style={draw=none},
  tick align=outside,
  xmajorgrids=true,
  grid style={gray!20,line width=0.3pt},
  bar width=4.5pt,
  bar shift=0pt,
  nodes near coords,
  nodes near coords align={horizontal},
  every node near coord/.append style={font=\scriptsize,text=black,xshift=1pt},
  clip=false
]
\addplot[fill=ttcblue!75,draw=none] coordinates {
  (859,1)
  (814,2)
  (793,3)
  (823,4)
  (871,5)
  (861,6)
  (821,7)
  (838,8)
  (952,9)
  (896,10)
};
\addplot[fill=gray!55,draw=none] coordinates {(794,11)};
\addplot[sharp plot,gray!85,dashed,line width=0.6pt,no marks,
  nodes near coords={},forget plot] coordinates {(794,0.55) (794,11.45)};
\end{axis}
\end{tikzpicture}
\vspace{-0.2cm}
\caption{Encounters with minimum TTC below 1.5~s. The dashed line marks the benign count.}
\vspace{-0.3cm}
\label{fig:ttc-encounters}
\vspace{-0.2cm}
\end{figure}
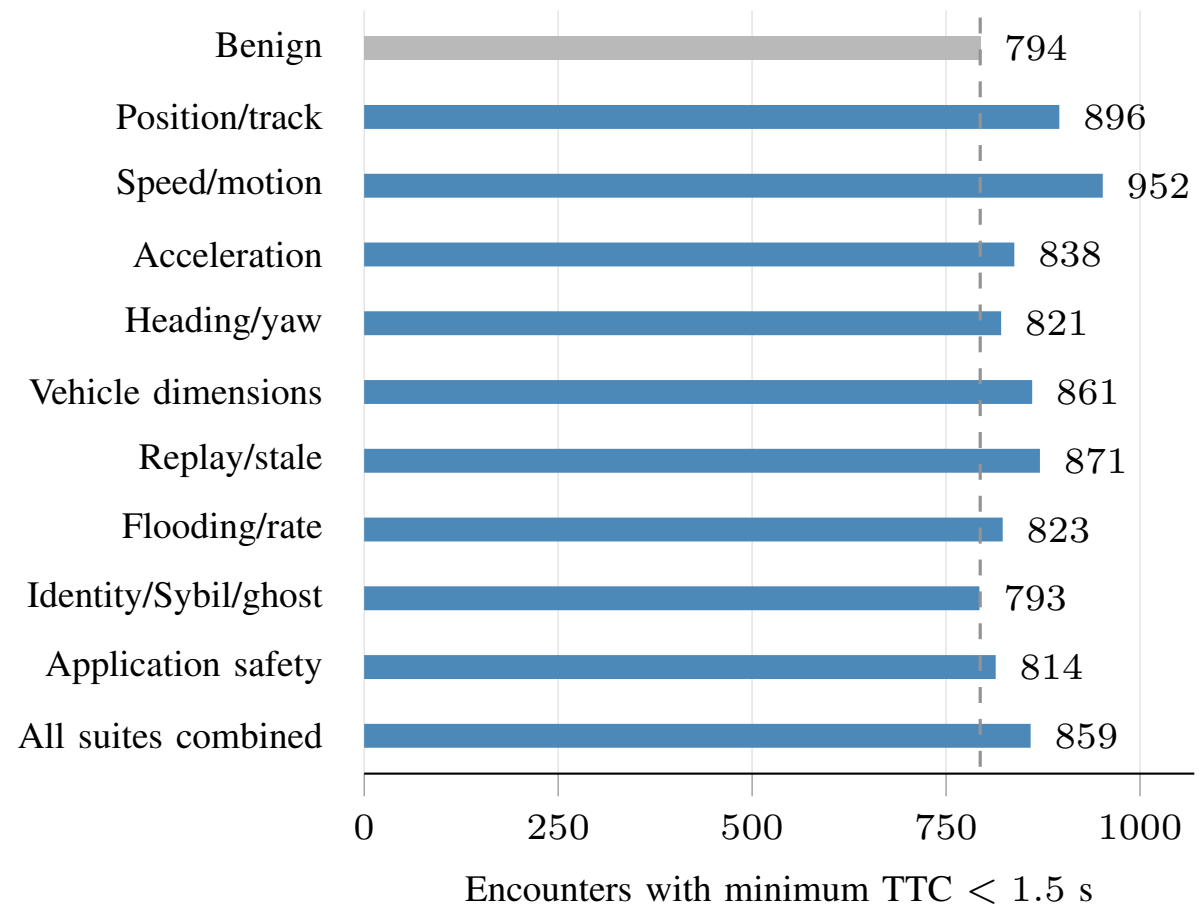

We next examined whether the received information was used to control a vehicle. In every attack case, the action arbiter selected speed requests linked to attacked messages. Table~\ref{tab:reaction-impact} reports these requests by application, the number of controlled vehicles receiving the resulting commands, and the recorded collision outcomes. The recorded execution times were approximately 3–4 hours per run. ICRW accounted for most selected requests in the speed and acceleration cases, with 3,750 and 646 requests, respectively. CAM-following instead accounted for most requests in the flooding, vehicle-dimensions, and identity cases. EEBL contributed selected requests linked to attacked messages only in the application-safety and combined cases, with five and two requests, respectively. These two cases contain the false emergency-braking profiles. The recorded trajectories show how vehicles reacted to the selected speed commands. In one \texttt{ConstSpeed} encounter, a stationary sender reported a speed of 46.19~m/s. At 402.21~s, the following vehicle’s CAM-following application used this information to request 18~m/s, which the arbiter selected. The vehicle accelerated from 0.57~m/s and collided with the stationary sender at 404.20~s; its recorded collision speed was 6.57~m/s. The transmission, reception, controller, and SUMO records link the false speed report to the selected command and the vehicle movement. Across the cases, the speed and longitudinal motion suite recorded the largest collision increase: 105 events compared with 66 in the benign case, an increase of 59.1\%. Flooding recorded the most attacker transmissions (234,637), but its collision total was 65. The highest transmission count, therefore, did not correspond to the largest collision increase in this campaign. Seventeen collisions in the speed case were attack-associated. The combined case had the same collision total as the benign case, although seven collisions were attack-associated. Total collision counts alone therefore do not show which vehicles acted on attacked information, and a lower total is not evidence of improved safety. Together, the message, command, and trajectory records show how a received attack can lead to a physical vehicle response. Collision totals alone can hide differences in traffic interactions. Fig.~\ref{fig:ttc-encounters} compares low-TTC encounters across all eleven cases. The vehicle-dimensions and combined cases both matched the benign total of 66 collisions. However, they recorded 861 and 859 low-TTC encounters, respectively, compared with 794 in the benign case, increases of 8.4\% and 8.2\%.

\begin{table*}[t]
\caption{Application use of attacked messages, recorded collision outcomes, and attacker transmissions. The CAM column refers to the CAM-following application. All cases use the same reaction-controlled vehicles. Attacker transmissions include unchanged messages. Attack-associated collisions are included in the total; association does not establish causation.}
\vspace{-0.2cm}
\label{tab:reaction-impact}
\centering
\footnotesize
\setlength{\tabcolsep}{3pt}
\begin{tabular}{@{}p{0.27\textwidth}cccccccc@{}}
\hline
& \multicolumn{3}{c}{\shortstack{\textbf{Selected}\\\textbf{attack-fed requests}}} & \shortstack{\textbf{Vehicles receiving}\\\textbf{attack-fed commands}} & \shortstack{\textbf{Total}\\\textbf{collisions}} & \shortstack{\textbf{Change from}\\\textbf{benign}} & \shortstack{\textbf{Attack-associated}\\\textbf{collisions}} & \shortstack{\textbf{Attacker}\\\textbf{transmissions}} \\
\textbf{Case} & \textbf{CAM} & \textbf{EEBL} & \textbf{ICRW} & & & & & \\
\hline
Benign & 0 & 0 & 0 & 0 & 66 & --- & 0 & 0 \\
Position and track presence & 17 & 0 & 16 & 11 & 74 & $+8$ & 1 & 118,718 \\
Speed and longitudinal motion & 175 & 0 & 3,750 & 46 & 105 & $+39$ & 17 & 117,599 \\
Acceleration & 479 & 0 & 646 & 52 & 62 & $-4$ & 3 & 116,840 \\
Heading and yaw rate & 189 & 0 & 32 & 34 & 75 & $+9$ & 0 & 117,429 \\
Vehicle dimensions & 1,082 & 0 & 112 & 49 & 66 & $0$ & 0 & 118,295 \\
Replay and stale data & 176 & 0 & 53 & 58 & 61 & $-5$ & 0 & 117,784 \\
Flooding and message rate & 1,240 & 0 & 76 & 69 & 65 & $-1$ & 0 & 234,637 \\
False identity, Sybil, and ghost vehicles & 1,015 & 0 & 58 & 48 & 76 & $+10$ & 7 & 157,145 \\
Application-targeted safety attacks & 114 & 5 & 92 & 20 & 60 & $-6$ & 1 & 119,070 \\
All suites combined & 348 & 2 & 164 & 42 & 66 & $0$ & 7 & 126,555 \\
\hline
\end{tabular}
\vspace{-0.3 cm}
\end{table*}

\vspace{-0.2cm}
\subsection{Indirect Vehicle Reactions}
\vspace{-0.1 cm}

\begin{figure}[t]
\centering
\definecolor{feedbackblue}{HTML}{1261A0}
\definecolor{feedbackred}{HTML}{A33B20}
\definecolor{feedbackgreen}{HTML}{176849}
\begin{tikzpicture}
\begin{groupplot}[
  group style={group size=1 by 2, vertical sep=30pt},
  scale only axis,
  width=0.82\columnwidth, height=0.29\columnwidth,
  xmin=1356.8, xmax=1359.4, ymin=-0.2, ymax=8.7,
  xtick={1357,1358,1359}, ytick={0,2,4,6,8},
  scaled x ticks=false,
  xticklabel style={/pgf/number format/fixed, /pgf/number format/1000 sep={}},
  tick label style={font=\scriptsize},
  label style={font=\scriptsize},
  ylabel={Speed (m/s)},
  axis lines=left, tick align=outside,
  ymajorgrids=true, grid style={gray!25, line width=0.3pt},
  clip=false
]
\nextgroupplot[
  xlabel={(a) Vehicle A: direct response},
  xlabel style={font=\footnotesize, yshift=-2pt},
  xticklabels=\empty,
  legend columns=2,
  legend style={at={(0.5,1.03)}, anchor=south, draw=none,
  font=\scriptsize, column sep=5pt, inner sep=1pt}
]
\addplot[feedbackblue, line width=1pt, const plot, mark=none] coordinates {
  (1356.8,0.00)
  (1356.9,0.00)
  (1357.0,0.00)
  (1357.1,0.00)
  (1357.2,0.00)
  (1357.3,0.00)
  (1357.4,0.26)
  (1357.5,0.52)
  (1357.6,0.78)
  (1357.7,1.04)
  (1357.8,1.16)
  (1357.9,1.16)
  (1358.0,1.16)
  (1358.1,1.16)
  (1358.2,1.16)
  (1358.3,1.16)
  (1358.4,1.42)
  (1358.5,1.68)
  (1358.6,1.94)
  (1358.7,2.20)
  (1358.8,2.46)
  (1358.9,2.72)
  (1359.0,2.98)
  (1359.1,3.24)
  (1359.2,3.50)
  (1359.3,3.76)
  (1359.4,3.31)
};
\addlegendentry{Recorded speed}
\addplot[gray!75, line width=0.65pt, dashed, const plot, mark=none] coordinates {
  (1356.800000000,0)
  (1357.364447601,1.1559703025554562)
  (1358.364447597,4.952092966954598)
  (1359.364447589,0)
  (1359.400000000,0)
};
\addlegendentry{Selected target}
\addplot[only marks, mark=*, mark size=1.5pt, feedbackred, forget plot] coordinates {
  (1357.364447601,1.1559703025554562)
  (1358.364447597,4.952092966954598)
};
\node[anchor=west, align=left, font=\scriptsize, inner sep=0pt]
  at (axis cs:1356.93,6.4) {Falsified acceleration\\received by A};
\draw[->, feedbackred, line width=0.5pt]
  (axis cs:1357.42,5.1) -- (axis cs:1357.364447601,1.1559703025554562);
\nextgroupplot[
  xlabel={Simulation time (s)\\[2pt](b) Vehicle B: subsequent response},
  xlabel style={font=\footnotesize, align=center}
]
\addplot[feedbackblue, line width=1pt, const plot, mark=none] coordinates {
  (1356.8,5.37)
  (1356.9,4.92)
  (1357.0,4.47)
  (1357.1,4.02)
  (1357.2,3.57)
  (1357.3,3.12)
  (1357.4,2.67)
  (1357.5,2.22)
  (1357.6,1.77)
  (1357.7,1.32)
  (1357.8,0.87)
  (1357.9,0.42)
  (1358.0,0.00)
  (1358.1,0.00)
  (1358.2,0.00)
  (1358.3,0.00)
  (1358.4,0.24)
  (1358.5,0.48)
  (1358.6,0.72)
  (1358.7,0.96)
  (1358.8,1.20)
  (1358.9,1.44)
  (1359.0,1.68)
  (1359.1,1.92)
  (1359.2,2.16)
  (1359.3,2.40)
  (1359.4,2.64)
};
\addplot[gray!75, line width=0.65pt, dashed, const plot, mark=none] coordinates {
  (1356.800000000,0)
  (1357.308847248,0)
  (1358.308847248,2.792904875155933)
  (1359.308847253,7.971654391254603)
  (1359.400000000,7.971654391254603)
};
\addplot[only marks, mark=*, mark size=1.5pt, feedbackgreen, forget plot] coordinates {
  (1358.308847248,2.792904875155933)
  (1359.308847253,7.971654391254603)
};
\node[anchor=west, align=left, font=\scriptsize, inner sep=0pt]
  at (axis cs:1356.93,6.4) {A's unmodified CAM\\received by B};
\draw[->, feedbackgreen, line width=0.5pt]
  (axis cs:1357.66,5.2) -- (axis cs:1358.308847248,2.792904875155933);
\end{groupplot}
\end{tikzpicture}
\caption{Direct and subsequent vehicle responses in the combined case. A reacts to falsified acceleration reports; B reacts to A's genuine movement reports. Solid lines show SUMO speed samples at 0.1-s intervals, and dashed lines show the selected speed commands. Coloured markers identify commands requesting acceleration.}
\vspace{-4pt}
\label{fig:live-feedback}
\vspace{-10pt}
\end{figure}
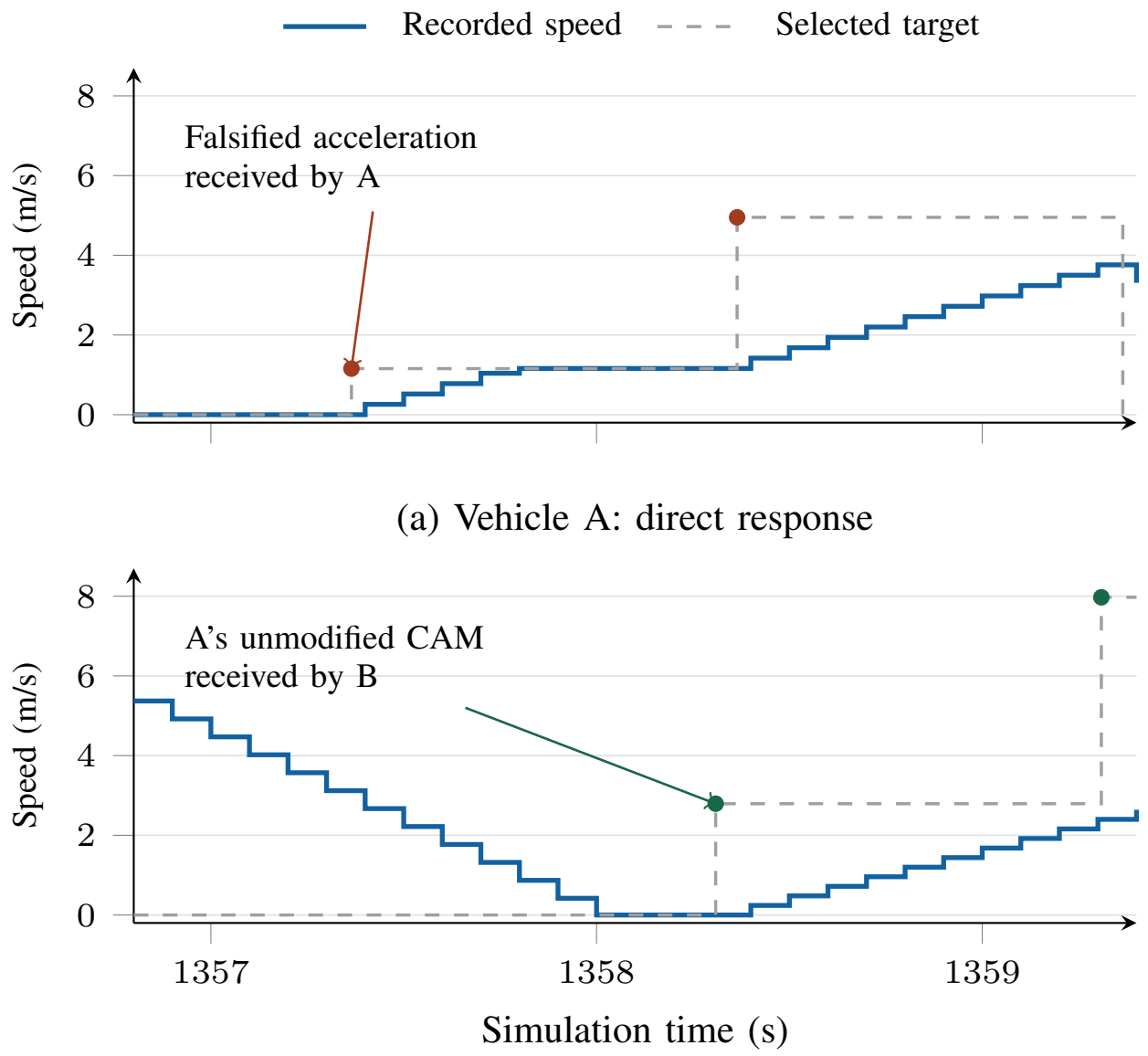

The effect of an attack can extend beyond the vehicle that acts on the falsified message. Fig.~\ref{fig:live-feedback} shows one such sequence from the combined-suite case. At 1,357.36~s, a stationary attacker reported an acceleration of 1.9~m/s$^2$. Vehicle~A used this CAM to request 1.16~m/s and began moving. Its next CAM reported its actual movement without any attack modification. Vehicle~B received that message at 1,358.31~s, requested 2.79~m/s, and accelerated from rest. Both receivers were non-attacking vehicles. During the plotted interval, B's selected commands came from A's unmodified CAMs rather than directly from attacked messages. This sequence showed how an attack can affect another vehicle through an unmodified message. A's reaction changed its movement, which it subsequently reported to B. Capturing this feedback required vehicle reactions to update the traffic simulation while it was running. By contrast, editing saved messages alone, as in VeReMi NextGen~\cite{vereminextgen2026}, changes message records but does not update the traffic simulation or generate subsequent messages reflecting new vehicle reactions.

\vspace{-0.1 cm}
\section{Conclusion}
\vspace{-0.1cm}
We extended VaN3Twin with 76 configurable attack profiles organized into nine suites, closed-loop vehicle reactions, and labelled dataset generation for benign and attack cases. Evaluation using the LuSTNano scenario verified the execution of all attack profiles and showed that every attack suite produced vehicle-control requests linked to attacked messages. The results also showed that collision totals alone do not capture the effects of attacked information, as some cases recorded more low-TTC encounters even when collision counts remained unchanged. Beyond direct vehicle reactions, we demonstrated how a vehicle responding to falsified information can subsequently influence another vehicle through unmodified messages reflecting its updated movement. By connecting attack injection, application decisions, vehicle reactions, and subsequent traffic interactions, the framework enables traceable evaluation of both direct and indirect attack effects as traffic evolves. The evaluation was limited to one traffic scenario with fixed random seeds. Future work will integrate misbehaviour detection, extend the attack library, and evaluate the framework across additional traffic scenarios and random seeds. The dataset's linked messages, decisions, and trajectories can also support future studies of how different attacks affect vehicle behaviour and traffic safety.
\vspace{-0.35cm}

\bibliographystyle{IEEEtran}
\bibliography{references}
\vspace{-0.2cm}

\end{document}